\documentclass{cimento}
\usepackage{graphicx} 

\title{Light Meson Spectroscopy with GlueX and Beyond}
\author{A.~Austregesilo\thanks{Contact: aaustreg@jlab.org} for the GlueX Collaboration}
\instlist{\inst{} Thomas Jefferson National Accelerator Facility, Newport News, VA, USA}

\begin{document}

\maketitle

\begin{abstract}
The GlueX experiment at Jefferson Lab was specifically designed for precision studies of the light-meson spectrum. For this purpose, a photon beam with energies up to 12\,GeV is directed onto a liquid hydrogen target contained within a hermetic detector with near-complete neutral and charged particle coverage. Linear polarization of the photon beam with a maximum around 9\,GeV provides additional information about the production process. In 2018, the experiment completed its first phase, recording data with a total integrated luminosity above 400\,pb$^{-1}$. We highlight a selection of results from this world-leading data set with emphasis on the search for light hybrid mesons. In the mean time, the detector underwent significant upgrades and is currently recording data with an even higher luminosity. The future plans of the GlueX experiment to explore the meson spectrum with unprecedented precision are summarized.
\end{abstract}

\section{Introduction}

The strong interaction is described by Quantum-Chromodynamics within the Standard Model of particle physics, but Confinement and the large coupling constant at low energies prevent the deduction of the hadron spectrum from first principles. For this reason, the precise experimental determination of the spectrum of hadrons is essential to further the understanding of the dynamics of the strong interaction.

The states in the light meson spectrum are characterized by their total angular momentum $J$, their parity $P$ and their charge conjugation $C$.  As they are broad and overlapping in mass, they have to be identified via the angular distribution of their decay products, and interference between resonances can be used to search for very small signals. Only certain $J^{PC}$ combinations are allowed for the mesons when described as quark-antiquark systems in the constituent quark model. In the recent past, several experiments~\cite{Dzi03,Noz09,Ale10, Rod19} have reported the observation of exotic states with forbidden quantum numbers, like the $\pi_1(1600)$ meson with $J^{PC}=1^{-+}$. However, their existence and interpretation is still debated.

\section{Light Meson Spectroscopy: The GlueX Experiment}

\begin{figure}
\centerline{\includegraphics[width=.7\textwidth]{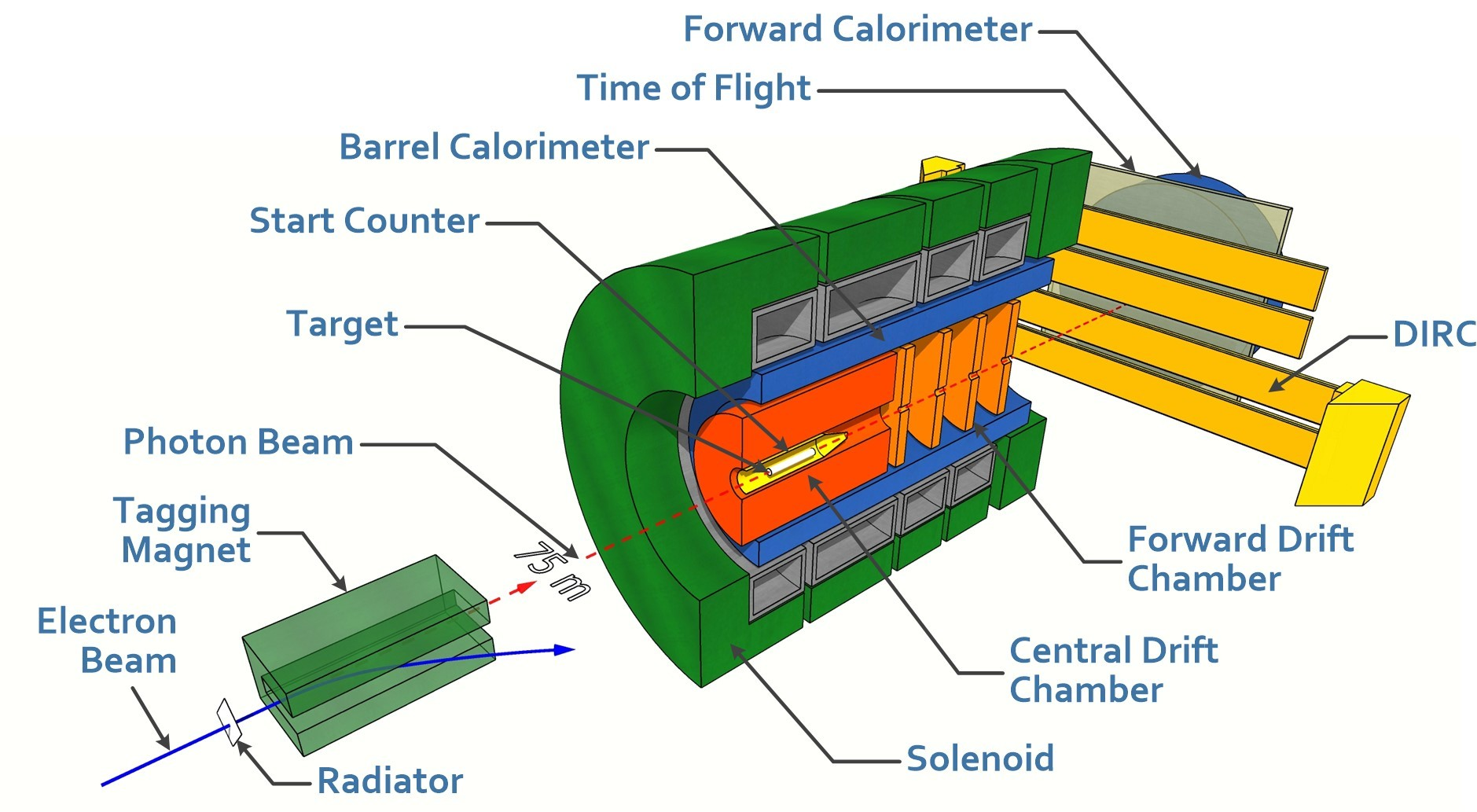}}
\caption{Schematic of the GlueX detector highlighting detector subsystems.\label{fig:det}}
\end{figure}

The GlueX experiment~\cite{Adh21} at the Thomas Jefferson National Accelerator Facility is part of the global effort to study the spectrum of hadrons. A primary electron beam of up to 12\, GeV is used to produce a secondary photon beam which impinges on a liquid-hydrogen target. Assuming vector meson dominance in $t$-channel production, a wide variety of mesonic states are accessible. A high beam intensity provides a sufficiently large reaction rate to study rare processes. The GlueX detector was specifically designed to map the light-quark meson spectrum up to masses of approximately 3\,GeV$/c^2$ with nearly complete acceptance for all decay modes. A superconducting solenoid magnet with a 2\,T field houses the target, central and forward drift chambers, and a barrel calorimeter~(see~Fig.~\ref{fig:det}). A forward calorimeter completes the forward photon acceptance and a time-of-flight counter provides particle identification capability.

The first phase of the GlueX experiment was completed in 2018, having recorded a total integrated luminosity above $400\,\mathrm{pb}^{-1}$. From 2019 onwards, a DIRC detector was added to the apparatus to improve the kaon identification. GlueX is scheduled to take data with an even higher luminosity until 2025 which will quadruple the existing data set.

The unique feature of GlueX is the capability to use a polarized photon beam. Polarization of the photons is achieved by coherent Bremsstrahlung of the primary electron beam on a thin diamond radiator. With a collimator suppressing the incoherent Bremsstrahlung spectrum, a linear polarization of up to 40\% is achieved in the coherent peak at 9\,GeV. In order to cancel apparatus effects, the polarization plane is rotated in steps of 45$^\circ$ around the beam axis into four different orientations during physics data taking. The degree of polarization is measured using the effect of triplet production~\cite{Dug17}. 

\subsection{Investigations of the Photoproduction Mechanism}

The photon beam polarization poses constraints on the quantum numbers of the produced meson systems. It may be used as a filter to enhance particular resonances or as additional input in multidimensional amplitude analyses. To this end, the photoproduction mechanism has to be understood in great detail, but only very limited data from previous experiments are available at these energies. GlueX has measured the beam asymmetry for the production of several pseudoscalar mesons: $\gamma p\rightarrow \pi^0 p$~\cite{Glx17}, $\gamma p\rightarrow \eta p$ and $\gamma p\rightarrow \eta^{\prime}(958) p$~\cite{Glx19}, $\gamma p \rightarrow  K^{+} \Sigma^{0}$~\cite{Glx20}, and $\gamma p \rightarrow \pi^- \Delta^{++}(1232)$~\cite{Glx21}. Many of these were the first measurements in this energy range. Common to all reactions, the natural parity exchange component dominates the production process for a single pseudoscalar meson. Only in charge-exchange reactions at small 4-momentum transfer $t$, contributions from unnatural exchanges, e.g. pion exchange, become relevant.

As an extension to this program, the measurement of spin-density matrix elements (SDMEs) quantify the transfer of the photon beam polarization to a system with spin. With the linearly-polarized beam, 9 real components of the complex-valued spin-density matrix can be accessed. When the production of a vector meson is described through diffractive scattering with $s$-channel helicity conservation, all but two of the SDMEs should be zero when measured in the helicity system. Our measurement of the SDMEs for $\rho(770)$ meson photoproduction~\cite{Glx23a} confirms this model at low values of 4-momentum transfer $t$, but observes significant deviation for larger $t$. In particular, the production process is sensitive to the interplay between Pomeron and $f_2/a_2$ exchanges, which can be modeled by Regge theory~\cite{Mat18}. We also determined the SDMEs for the process $\gamma p \rightarrow K^+\Lambda(1520)$~\cite{Glx22} and are extending this program to the vector mesons $\phi$, $\omega$ and even $J/\psi$~\cite{Chu23}. The results of these measurements will improve the understanding of the photoproduction process. In addition, these studies are used to evaluate the accurate modelling of the GlueX detector acceptance and study systematic effects.

\subsection{Amplitude Analysis of Multi-Meson Final States}

Whenever several overlapping meson resonances decay into the same final state, we use the technique of amplitude analysis to disentangle states via the angular distribution of their decay products. Complex-valued amplitudes describe the dynamics of the decay, and an extended maximum-likelihood fit is used to extract the production strengths.

As the strongest evidence for an exotic $J^{PC}=1^{-+}$ amplitude was reported in $\eta^{(')}\pi$~\cite{Rod19}, we focused our initial effort on these final states. The GlueX data reaches competitive statistical precision to previous experiments, is using a complementary production mechanism and has access to different production and multiple decay modes. In order to use the photon beam polarization to separate natural from un-natural parity exchange processes, a new amplitude formalism had to be developed in collaboration with the Joint Physics Analysis Center (JPAC)~\cite{Mat19}. This formalism is used to describe all systems with two pseudoscalar mesons, and an extension to vector-pseudoscalar combinations has since been developed~\cite{Sch23}.

The complexity of this new amplitude formalism impede a fully mass-independent analysis so far, but the analysis is stabilized by imposing a Breit-Wigner resonance shape for the dominant $a_2(1320)$ signal. Extracting the cross section for the $a_2(1320)$ photoproduction~\cite{Alb23} yields a good agreement with theory predictions and thus demonstrates the validity of this method. The results are used as a reference for the search for the exotic $\pi_1(1600)$ in this channel. In addition, branching fractions from recent Lattice QCD computations~\cite{Wos21} are combined with measurements of the iso-vector $b_1(1285)\pi$ cross section to determine upper limits for $\pi_1$ production. While a small signal is expected in $\eta\pi$, the signal could be dominant in $\eta'\pi$ where we place an additional effort.

\subsection{GlueX Phase 2 and the JLab Eta Factory}

For the remaining beam time in 2024/25, the GlueX forward calorimeter is currently being upgraded with a PbWO$_4$ insert, which provides higher granularity and improved radiation hardness in the central part. The GlueX program will continue to take data with high luminosity, focusing on final states with kaons to map the strangeness component of the light mesons spectrum. In parallel, the JLab Eta Factory (JEF) aims to perform precision measurements of various $\eta$ decays with emphasis on rare neutral modes. Since the production rates of $\eta$ and $\eta'$ are similar under these experimental conditions, the same data set will also offer sensitive probes for $\eta'$ decays. Compared to all existing and planned $\eta/\eta'$ experiments in the world, the unique feature of the JEF program is a clean data set in the rare neutral decays of $\eta$ and $\eta'$ with up to two orders of magnitude background reduction. 

\section{Probing the Electromagnetic Structure of Hadrons}

The versatile GlueX experimental setup is also used to probe the electromagnetic structure of hadrons. In the Primakoff process, beam photons scatter on the strong electromagnetic field that surrounds the nucleus in a fixed target geometry.

\subsection{PrimEx-$\eta$: Measurement of the $\eta\rightarrow\gamma\gamma$ decay width}

This experiment aims for a precision measurement of the $\eta \rightarrow \gamma\gamma$ decay width, which will be extracted from the measured differential cross sections at forward angles on light targets, $^4$He and Be, using a tagged photon beam with energies up to 11.5\,GeV. The result will not only potentially resolve a long standing discrepancy between the Primakoff and collider measurements, but is expected to reduce the experimental uncertainty of the current PDG average~\cite{Wor22} on this important quantity by a factor of two. This will result in a direct improvement on all other partial decay widths of the $\eta$ meson. The high precision measurement will have significant impact on the experimental determination of the fundamental parameters, such as the ratios of light quark masses and the $\eta - \eta'$ mixing angle. The experiment was completed in three beam times from 2019 to 2022 and the analysis is ongoing. A precise measurement of the Compton cross section over the full energy range is currently used to evaluate the systematic uncertainties of the experiment.

\subsection{Charged and Neutral Pion Polarizability}

The polarizability describes the deformation of an object when subjected to an external electric field. For pions, this fundamental property of the strong interaction has precise predictions from chiral perturbation theory and is expected to be extremely small. Only the strong electromagnetic field produced by nuclei is able to deform the hadrons measurably. In our experiment, we scattered a 6\,GeV polarized photon beam on a lead target to produce pion pairs, $\pi^+\pi^-$ and $\pi^0\pi^0$, and determine the cross section for these processes. A new wire chamber was installed downstream of the forward calorimeter to be able to detect muon pairs from Bethe-Heitler production, which is a background process but can be used to normalize the cross section measurement for the charged pion polarizability. The experiment was completed in 2022 and a competitive precision with previous state-of-the art experiments is expected. The inverted kinematics from traditional Primakoff experiments~\cite{Ado15} and the photon beam polarization are the unique features of this complementary measurement of the charged pion polarizability. The polarizability for the neutral pions has never been measured before.

\section{Summary and Outlook}

We presented an overview of the light meson spectroscopy program currently pursued at the GlueX experiment at Jefferson Lab. The full data set recorded in the initial phase of the experiment is available and under active analysis, with several exciting results already published and presented at this conference~\cite{Chu23,Sch23,Alb23,Ner23}. Several necessary milestone towards the search for exotic mesons have been reached, and the mapping of the light meson  spectrum is well underway. Close collaboration with theory on model development and interpretation is indispensable to reach this ambitious goal.

In parallel, the GlueX detector was equipped with improved particle identification and calorimetry, and the data taking for the second phase is in process. By the end of 2025, we expect to quadruple the existing data set which will allow us to study rare processes and explore the strangeness dimension of the light meson spectrum. Besides this program, the versatile GlueX detector is also used for precision measurements of the electromagnetic structure of hadrons.

After the GlueX program is complete, we plan to transform the photon beam line into a tertiary beam of neutral long-lived kaons. This world-wide unique $K_\mathrm{L}$ beam facility combined with the multi-purpose GlueX detector is planned to be ready after 2026. It will open up the possibility to search for missing hyperon resonances in the baryon spectroscopy sector and allow us to study the $\kappa$ meson in $K\pi$ scattering.

Furthermore, we are currently developing a program for hadron spectroscopy opportunities at a possible 22\,GeV electron accelerator upgrade at Jefferson Lab. This beam energy would span far beyond the threshold for charmonium production, and would allow us to search for the photoproduction of the recently discovered $X$, $Y$, $Z$ and $P_c$ states at GlueX.

\acknowledgments
Work supported by the U.S. Department of Energy, Office of Science, Office of Nuclear Physics under contract DE-AC05-06OR23177. The author acknowledges the support by the DOE, Office of Science, Office of Nuclear Physics in the Early Career Program.


\begin{thebibliography}{0}
\bibitem{Dzi03} \BY{A. R. Dzierba et al. (E852)}
    \IN{Phys. Rev. D}{67}{2003}{094015};
\bibitem{Noz09} \BY{M. Nozar et al. (CLAS)}
    \IN{Phys. Rev. Lett.}{102}{2009}{102002};
\bibitem{Ale10} \BY{M. G. Alekseev et al. (COMPASS)}
    \IN{Phys. Rev. Lett.}{104}{2010}{241803};
\bibitem{Rod19} \BY{A. Rodas et al. (JPAC)}
    \IN{Phys. Rev. Lett}{122}{2019}{042002};
\bibitem{Adh21} \BY{S. Adhikari et al. (GlueX)}
    \IN{Nucl. Instrum. and Meth. A}{987}{2021}{164807};
\bibitem{Dug17} \BY{M. Dugger et al.}
    \IN{Nucl. Instrum. and Meth. A}{867}{2017}{115};
\bibitem{Glx17} \BY{H. A. Ghoul et al. (GlueX)}
    \IN{Phys. Rev. C}{95}{2017}{042201};
\bibitem{Glx19} \BY{S. Adhikari et al. (GlueX)}
    \IN{Phys. Rev. C}{100}{2019}{052201(R)};
\bibitem{Glx20} \BY{S. Adhikari et al. (GlueX)}
    \IN{Phys. Rev. C}{101}{2020}{065206};
\bibitem{Glx21} \BY{S. Adhikari et al. (GlueX)}
    \IN{Phys. Rev. C}{103}{2021}{L02201};
\bibitem{Glx23a} \BY{S. Adhikari et al. (GlueX)}
    \IN{accepted by Phys. Rev. C}{}{2023}{};
\bibitem{Mat18} \BY{V. Mahieu et al. (JPAC)}
    \IN{Phys. Rev. D}{97}{2018}{094003};
\bibitem{Glx22} \BY{S. Adhikari et al. (GlueX)}
    \IN{Phys. Rev. C}{105}{2022}{035201};
\bibitem{Chu23} \BY{E. Chudakov (GlueX)}
    \IN{these proceedings}{}{2023}{};
\bibitem{Mat19} \BY{V. Mathieu et al. (JPAC)}
    \IN{Phys. Rev. D}{100}{2019}{054017};
\bibitem{Sch23} \BY{A. M. Schertz (GlueX)}
    \IN{these proceedings}{}{2023}{};
\bibitem{Alb23} \BY{M. Albrecht (GlueX)}
    \IN{these proceedings}{}{2023}{};
\bibitem{Wos21} \BY{A. J. Woss et al. (HadSpec)}
    \IN{Phys. Rev. D}{103}{2021}{054502};
\bibitem{Wor22} \BY{R. L. Workman et al. (PDG)}
    \IN{Prog. Theor. Exp. Phys.}{2022}{2022}{083C01};
\bibitem{Ado15} \BY{C. Adolph (COMPASS)}
 	  \IN{Phys. Rev. Lett.}{114}{2015}{062002};
\bibitem{Ner23} \BY{F. Nerling (GlueX)}
    \IN{these proceedings}{}{2023}{};
\end{thebibliography}
\end{document}